# RFID et nouvelles technologies de communication; enjeux économiques incontournables et problèmes d'éthique.

**André Thomas**

*Centre de Recherche en Automatique de Nancy*
*ENSTIB – B.P. 1041*
*88051 EPINAL Cedex. France*

*Andre.Thomas@cran.uhp-nancy.fr*

RÉSUMÉ :

Actuellement, le code à barres est progressivement remplacé par des puces RFID (pour Radio-Frequency Identification). Des apports dans des domaines tels que la traçabilité, la lutte contre les « contre-façons », le pilotage de la production, la domotique, etc… ont été démontrés. Mais on a vu même des puces sous-cutanées qui peuvent identifier des humains. Il y a donc des questions d'éthique sérieuses quant à leur application dans le monde actuel.

L'article s'attachera donc, dans une première partie, à présenter des éléments de techniques de la RFID, puis s'appuyant sur une description globale des applications actuelles, il en dégagera des apports et des perspectives intéressantes pour le monde économique. Enfin la deuxième partie, tentera de poser les risques relatifs aux individus et à la société en général, pour ensuite poser quelques questionnements relatifs aux problèmes d'éthique nécessitant des normalisations et réglementations.

MOTS-CLÉS : RFID, Identification, Produits intelligents, Environnements ambiants

**INTRODUCTION**

Actuellement, le code à barres est progressivement remplacé par des puces RFID (pour Radio-Frequency Identification) ou « étiquettes» communicantes ayant la possibilité de stocker de l'information de manière dynamique et de communiquer sans fil avec leur environnement ambiant. Or depuis quelques années, bon nombre de chercheurs et industriels ont montré que l'on pouvait tirer parti de cette « instrumentation des produits » via ces technologies. Des apports dans des domaines tels que la traçabilité, la lutte contre les « contre-façons », le pilotage de la production, la domotique, etc… ont été démontrés.

La prolifération de ces puces semble aujourd'hui très importante. On cherche à les attacher à des produits usuels, tels les passeports pour augmenter la sécurité, par exemple. Il suffit de présenter un « passe » à une borne de métro pour valider (et compter) ses trajets. Il n'est presque plus nécessaire de patienter pour valider son identité dans bon nombre d'endroits. On a vu même des puces sous-cutanées qui peuvent identifier des humains. Il y a donc des questions d'éthique sérieuses quant à leur application dans le monde actuel. Techniquement, on peut lire désormais les puces incorporées dans des vêtements et produits ordinaires sans que celui qui porte avec lui ces puces le sente. Serait-il possible de « traquer » des gens grâce à cette technologie ?!

L'article s'attachera donc, dans une première partie, à présenter des éléments de techniques de la RFID, puis s'appuyant sur une description globale des applications actuelles, il en dégagera des apports et des perspectives intéressantes pour le monde économique. Enfin la deuxième partie, tentera de poser les risques relatifs aux individus et à la société en général, pour ensuite poser quelques questionnements relatifs aux problèmes d'éthique nécessitant des normalisations et réglementations.

**LA TECHNOLOGIE**

La première expérience de puces radio pourrait être en 1939 lorsque l'armée britannique équipa ses avions d'un système (IFF pour Identification Friend or Foe). Sans vouloir faire toute l'histoire de cette technologie ni la décrire précisément car aujourd'hui beaucoup d'ouvrages ou d'articles l'ont déjà fait, nous pouvons juste faire remarquer que nous avons évolué vers plus de miniaturisation et de distance de lecture/écriture. Aujourd'hui ces technologies se décrivent par les fréquences et les performances qui les caractérisent. Par ailleurs et pour des raisons évidentes d'interopérabilité, nous avons aussi évolué vers plus de normalisation, à l'initiative de l'Auto-ID center du MIT (1999) qui préconise le standard EPC Global (créé par l'Uniform Code Council et EAN International).

**1/ Dans quelles bandes de fréquence fonctionnent les étiquettes électroniques ?**

Différentes bandes de fréquences aujourd'hui existent couramment. Pour certaines d'entre elles, les fréquences disponibles varient selon les régions du monde. De manière très schématique (cf. tableau 1 ci-dessous) on distingue trois grandes zones (la réalité est un peu plus compliquée, notamment en Asie et Océanie).

*Tableau 1 - répartition des fréquences selon la zone géographique*

| Fréquences | Europe et Afrique | Amériques Nord et Sud | Asie et Océanie |
|---|---|---|---|
| BF (125-135 kHz) *ISO 18000-2* | 125 kHz | 125 kHz | 125 kHz |
| HF (13,56 MHz) *ISO 18000-3* | 13,56 MHz | 13,56 MHz | 13,56 MHz |
| UHF (860-960 MHz) *ISO 18000-6* | 865 - 868 MHz ( 2W ERP (1) - LBT) | 902 - 928 MHz (4W - EIRP) | 902 - 928 MHz (2) (Japon 952 - 954 MHz) |
| SHF (2,45 GHz) *ISO 18000-4* | 2,446 - 2,454 GHz | 2,427 - 2,47 GHz | 2,4 - 2,4835 GHz |

1) 1W ERP (Effective Radiated Power) = 1,62W EIRP (Equivalent Isotropic Radiated Power)
2) Pas d'harmonisation. Par exemple : 952-954 MHz au Japon, 910-914 MHz en Corée

**2/ Quelles sont les performances des étiquettes électroniques ?**

On distingue en particulier :

- la rapidité de lecture (de manière générale, elle augmente avec la fréquence) ;
- la possibilité de lire plusieurs étiquettes de manière quasi-simultanée (par exemple lors d'un passage sur tapis roulant) ;
- la distance de lecture.

La distance de lecture d'une étiquette varie selon plusieurs paramètres :

- la bande de fréquences utilisée ;
- les paramètres réglementaires relatifs à cette bande de fréquences, en particulier :
  puissance du lecteur ;
  modulation et protocole d'émission.
- les performances des composants silicium utilisés et la sensibilité du lecteur ;
- le recours éventuel à une batterie (qui peut porter cette distance à 100 m) ;

- l'environnement d'utilisation et le déploiement effectué (niveau de bruit électromagnétique ; présence d'humidité ; matériaux de l'objet étiqueté ; réflexions multiples ; interférences entre lecteurs, géométrie, etc...).

Chaque bande de fréquences a ses avantages et ses inconvénients par rapport au contexte et aux finalités d'utilisation. Les principales caractéristiques liées aux bandes de fréquences sont présentées dans le tableau 2 ci-dessous.

*Tableau 2 - principales caractéristiques liées aux bandes de fréquences*

|  | BF | HF | UHF | 2,45 GHz |
|---|---|---|---|---|
| Sensibilité à l'eau, l'humidité |  |  | Relativement sensible de préférence produits secs / atmosphère sèche | Forte sensibilité |
| Environnement métallique |  | déploiement susceptible d'être plus difficile | Potentiellement néfaste, s'il n'est pas géré |  |
| Caractéristiques des champs | Couplage inductif (*) | Couplage inductif * | Propagation d'onde très sensible aux matériaux de l'environnement (**) |  |
| Débit | Quelques kbits/s | De l'ordre de 100 kb/s | De l'ordre de quelques centaines de kbits/s |  |
| Portée type pratique(***) | De l'ordre du m ; pénètre légèrement le métal | De l'ordre du m | De l'ordre de 4-5 m (Europe, US) |  |

\* les caractéristiques du champ électromagnétique sont relativement maîtrisables.
\*\* sensible aux réflexions et à l'absorption - les caractéristiques du champ sont moins faciles à maîtriser.
\*\*\* fortement liée à la réglementation applicable.

Ainsi ces tags peuvent être passifs ou actifs, en lecture seule ou en lecture/écriture, à positionnement discret ou géolocalisés, de taille de quelques centimètres ou de quelques centièmes de millimètres, etc…

**LES APPORTS**

Ces technologies proliférant, nous sommes en train de passer à une échelle complètement différente au regard de la communication : nous entrons dans l'ère de « l'internet des objets ». Le réseau mondial (la toile) des ordinateurs n'est désormais plus suffisant, chaque objet, chaque personne portant un tag pourrait en faire partie ! Le système est prêt, offrant un potentiel d'exploitation impressionnant. Dans ce contexte, deux sphères sont concernées : celles de la vie professionnelle et celle de la vie privée. C'est le concept d'ubiquité de la connectivité dans le temps et l'espace.

Il est évident qu'une multitude d'usages peut être imaginé dès lors que les objets portent de l'information, peuvent en capter, peuvent en distribuer à d'autres de leur entourage, voire peuvent participer à des décisions localement…

Un produit pourrait tout dire de lui… Nous imaginons encore très difficilement les conséquences d'une telle fonction. Cela conduirait à mettre à disposition des milliards de milliards de données, ce qui est à l'heure actuelle encore une des plus grosses faiblesses, mais qui ouvre la possibilité de concevoir des « espaces intelligents ». On parle aujourd'hui « d'intelligence ambiante ».

Cette technologie peut améliorer le confort sous diverses formes. L'instrumentation d'une maison pourrait ainsi permettre d'optimiser la consommation d'énergie et la température ambiante en fonction des mouvements, des actions observés dans les pièces, par exemple. La sécurité des biens et des personnes pourrait aussi être améliorée. Il en est de même de la santé des individus, ne serait-ce que une meilleure gestion des informations qui leur sont propres.

Un système d'étiquettes RFID collées sur des kanbans dans une entreprise de l'ameublement a été prototypé et a pu montrer qu'en assistant l'opérateur dans ses décisions de gestion des priorités entre produits à réaliser sur un même poste de charge, il a conduit à des gains en productivité, fluidité et niveaux de stocks très significatifs.

## LES RISQUES

Mais si bon nombre de bienfaits sont déjà visibles et sont aussi prévisibles, nous pouvons mettre en évidence déjà de sérieux risques tant pour la santé que pour la liberté individuelle ou collective.
- L'impact de la prolifération des ondes radio n'a pas encore été évalué sur la santé.
- Les infrastructures mises en place permettent de surveiller les mouvements, voire les activités, des « porteurs » de tags. En effet, (le téléphone portable en est un bon exemple) au-delà des mouvements, la surveillance de l'évolution des informations contenues dans les tags permet d'avoir une trace et un suivi des activités qui les ont générées.
- Numéroter des objets et des personnes conduira à faire évoluer l'interprétation de l'identité et de l'individualité, conduisant ainsi à une évolution des relations sociales, à une modification des règles de la vie communautaire et même à une minimisation de la sphère d'intimité.
Par ailleurs, des menaces diverses pourraient survenir :
- Menaces d'actions inopportunes : dès observation, via des tags, que plusieurs objets sont en train de sortir simultanément d'un magasin, mise en fonctionnement d'un système anti-vol (vidéo ou autre) qui pourrait être « brutal » ou mettre les acteurs en situation délicate, alors le fait est normal.
- Menaces d'exploitation commerciale : l'identification personnelle des clients associée à la surveillance de leurs consommations dans un magasin pourraient permettre à un logiciel de CRM de « forcer » des publicités ou incitations à l'achat à l'insu de la clientèle.

- Menaces de détermination des goûts : concept identique au précédemment.
- Menaces de pertes de discrétion sur l'identité et localisation : Si on multiplie les tags et les lecteurs et que l'on suit en permanence et de manière géolocalisée les déplacements, on peut reconstruire des trajets, des arrêts réguliers et finir par localiser adresse, identité, habitudes, etc…des personnes (ce système fait déjà partie de projets de recherche pour des applications sur des trajets de produits). En élargissant, les personnes rencontrées pourraient de même être déterminées et ainsi le comportement, les actions des personnes seraient déduites.
- Menaces d'effets indirects : Si les données personnelles sont connues dans une base de données, elles pourraient être modifiées et alors si quelqu'un d'autre y accède, des informations fausses des personnes pourraient être diffusées.
Tout ceci n'est pas exhaustif, mais n'est pas non plus irréaliste.

Face aux diverses volontés des gouvernements, dans certaines circonstances, des dangers nouveaux pourraient effectivement apparaître. Par exemple, le Japon s'est donné comme objectif pour 2010 de devenir la « société du réseau ubique », l'idée est d'évoluer vers du contrôle (au sens de l'automatique) temps réel de plus en plus de choses grâce à la prolifération de dizaines et de dizaines de milliards de puces. Diverses applications ont déjà donné de bons résultats (gestion de certains animaux, des médicaments dans les hôpitaux, etc…). Cette évolution se fera naturellement par des phases différentes permettant la vulgarisation, le retour sur investissements, la crédibilité… On peut citer :

- La traçabilité et le suivi d'activité
- Le partage d'information (entre lieux, entre sociétés, entre systèmes)
- L'auto-organisation (vers le produit intelligent, vers les systèmes ambiants)

Nous allons vers un monde où les choses vont être animées … Mais comme dit Michel Alberganti dans son livre, animées « de quelles intentions » ?!

Paradoxalement, les personnes ne sont actuellement pas réellement contre. Une enquête réalisée en 2005 a montré que les personnes interrogées étaient plutôt favorables à la traçabilité car elle permettait plus de contrôle (92%), plus de sécurité (91%), plus de qualité (90%), plus de confiance (89%), plus de responsabilité (83%), plus de ré »activité (72%) …

## LES PROBLEMES D'ETHIQUE

La loi informatique et libertés reconnaît aux citoyens des droits spécifiques pour préserver leur vie privée et leurs libertés dans un monde numérique. La CNIL (Commission Nationale Informatique et Liberté) a été créée en 1978 pour aider les personnes à contrôler la bonne application de cette loi. Le système de protection est là… Mais à l'échelle de cette prolifération en aura-t-il les moyens ?

L'interprétation des messages et/ou informations échangés entre individus n'est pas nouvelle chose. Aujourd'hui, une tranche infime de la population en a le pouvoir (police, gouvernement, …). Avec la généralisation de la RFID, il n'en sera plus de même, les sociétés, les privés, toute personne ayant les moyens de lire, de décrypter des informations pourra faire de même. Le plus notable est encore de savoir qu'elles pourront le faire à l'insu de tous. Tout sera donc sous contrôle … de ceux qui en auront les moyens financiers et/ou techniques ! On peut même imaginer qu'ils pourront transgresser les règles inscrites dans le système global d'information. Seuls, les citoyens « ordinaires » n'auront plus qu'à les suivre.

Il nous paraît intéressant de classer les effets pervers d'une évolution mal contrôlée de cette technologie (ou de toute autre relative à l'identification), selon les critères suivants :

- Automatisation.

L'installation dans un quelconque endroit (entreprise, lieu public, …) d'un ou de plusieurs lecteurs conduit à la mise sous contrôle permanent et passif des objets et personnes tagés. Un lecteur peut ainsi contrôler toute une zone.

- Identification

Au passage sous le lecteur ou dans le champ de lecture pour les technologies de géolocalisation, il est possible d'identifier de manière statique, voire évolutive, les entités. Les porteurs de tags ont ainsi « l'obligation de collaborer », dans la mesure où des informations peuvent être « forcées » sur les tags à l'insu des porteurs.

- Intégration

Les tags peuvent aujourd'hui être plus ou moins intégrés aux porteurs (cousus dans un col de chemise, collé sur un objet, voire intégré dans la matière !). Ils sont donc plus ou moins visibles, plus ou moins à la connaissance des porteurs.

- Authentification

Les tags peuvent alors servir à l'authentification, via des informations privées, des porteurs, ceci pour un usage de gestion de la santé, pénal ou de contrôle de la vie privée.

Ainsi la lecture de ces tags peut induire des problèmes d'interrogation clandestine s'il n'y a pas consentement de l'individu (voire même de connaissance du fait). On peut imaginer qu'au passage à proximité d'un panneau publicitaire, une propagande apparaisse en fonction de la catégorie socioprofessionnelle du passant !

Le fait d'être aussi sous contrôle, ou « sur écoute », en permanence relèverait plus de l'espionnage que du contrôle de traçabilité fonctionnelle.

Enfin les informations véhiculées pourraient être captées par d'autres acteurs que ceux pour lesquelles elles ont été initialement mises en place et ainsi, dans un monde ubiquitaire, se retrouver dans les mains de personnes peu scrupuleuses. De la même manière, des informations non souhaitées pourraient être mises sur les tags et ainsi se déployer dans un système sans la volonté du porteur.

On peut ainsi imaginer des scénarii qui « justifieraient » de tels agissements aux yeux de certains, pour l'exemple citons en laissant le lecteur les imaginer :
- Scénario de propagande
- Scénario criminel
- Scénario de surveillance dans un contexte de crise
- Scénario de guerre

**CONCLUSION**

Il est ainsi de nos jours évident que si les technologies d'identification, et en particulier les technologies RFID, vont se déployer de manière extrêmement importante et apporteront beaucoup de gains en productivité, en sécurité, en réactivité par une meilleure traçabilité, il nous faut rester vigilant sur les dangers qui peuvent l'accompagner.

Le phénomène n'est pas exceptionnel. Toute modernisation, toute innovation est toujours sujette à ce genre de risque. La particularité de cette technologie est la dimension du déploiement qui lui est inhérent, d'une part, et les domaines mêmes qui seront concernés, d'autre part.

A ce jour, ces derniers ne nous sont pas encore entièrement connus. Et c'est une des raisons pour lesquelles nous en avons peur. Il nous faut donc être vigilants et la meilleure manière de l'être est d'en étudier toutes les possibilités afin de les rendre publiques et à la portée du maximum de personnes.

**ANNEXE**

> Qui surveille la RFID ?

- CNIL (Commission Nationale de l'Informations et des Libertés) 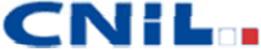
- CGTI (Conseil Général des Technologies de l'Information) 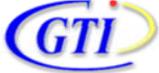
- Parlement européen 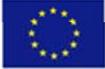

> Les lois relatives à la protection des données

- loi n° 78-17 du 6 janvier 1978 relative à l'informatique, aux fichiers et aux libertés modifié à divers reprises.
- Directive 2002/58/CE du Parlement européen et du Conseil concernant le traitement des données à caractère personnel et la protection de la vie privée dans le secteur des communications électroniques (JOCE du 31 juillet 2002).
- Charte européenne des droits fondamentaux, art. 8 (futur II-8 du projet de Constitution), adoptée par le Sommet européen de Nice, juin 2002.

**BIBLIOGRAPHIE**